# Raman Quantum Memory of Photonic Polarised Entanglement

**Dong-Sheng Ding**[†,1,2], **Wei Zhang**[†,1,2], **Zhi-Yuan Zhou**[1,2], **Shuai Shi**[1,2],

**Bao-Sen Shi**[1,2,*] **and Guang-Can Guo**[1,2]

[1]*Key Laboratory of Quantum Information, University of Science and Technology of China, Hefei,*

*Anhui 230026, China and*

[2]*Synergetic Innovation Center of Quantum Information & Quantum Physics, University of Science*

*and Technology of China, Hefei, Anhui 230026, China*

[†]*These authors contribute this article equally*

*\*Corresponding author: drshi@ustc.edu.cn*

Quantum entanglement of particles is regarded as a fundamental character in quantum information, in which quantum state should be given for whole system instead of independently describing single particle. Constructing quantum memory of photonic entanglement is essential for realizing quantum networks, which had been performed previously by many memory protocols. Of which Raman quantum memory gives advantages in broadband and high-speed properties, resulting in huge potential in quantum network and quantum computation. However, Raman quantum memory of photonic polarised entanglement is a challenge work and still missing. Here, we report two Raman quantum memories based on gas atomic ensembles: 1. Heralded Raman quantum memory of hybrid entanglement of single photon's path and polarization. 2. Raman storage of two-particle photonic polarised entangled state. Our experimental performances of these two different Raman quantum storages of photonic entanglement show a very promising prospective in quantum information science.

The most convenient and robust physical system for encoding quantum information is photons, which could be encoded with the superposition of its intrinsic degree of freedom e.g polarization, orbital angular momentum (OAM), path, time-bin, etc. Toward quantum communications in the future, quantum memory combined with photonic entanglement swapping could overcome the scaling of error rate exponentially increasing with the channel length [1]. Thus, efficient and controllable quantum storage of photonic entanglement is a significant step to synchronously establish quantum memories entanglement distribution over long-distance between quantum nodes [2-4].

There are many progresses in quantum memory of photonic entanglement by using atomic ensembles [5-9]. The techniques for realizing such quantum memory include electromagnetically induced transparency [10-12], far off-resonant two-photon transition [13-16], controlled reversible inhomogeneous broadening [17-20], atomic frequency combs [21,5,8], photon echo [22-23], optomechanical storage [24] and off-resonant Faraday [25-26]. The far off-resonant two-photon transition is a protocol by using a Raman excitation configuration, thus this storage is called as Raman quantum memory in below. Raman quantum memory takes advantages of off-resonance configuration and there is an excited virtual energy level of atoms near two-photon resonance, the band of the excited virtual energy level could be increased by enlarging the effective optical depth (OD) or increasing the intensity of control laser. Thus, this storage protocol could have the ability of storing short time pulse, whereas corresponding to a broad-band quantum memory. Besides, theoretically, Raman quantum memory could store large frequency range because the single photon's detuning of control laser is changeable. Therefore, by using such way of quantum memory, the speed of quantum computation in practical is enhanced and the establishment of quantum memories distribution would be efficient.

The Raman quantum memory protocol had been experimentally realized by Walmsley's group, in which the storage of nanosecond timing pulse width was performed in Refs [14-15]. This had been a significant progress in realization of Raman quantum memory despite the fact that the information carrier is the attenuated weak coherent light. Recently, our group and Walmsley's group reported the Raman quantum memory of true single photon [16, 27]. In our experiment, the two-particle photonic OAM entanglement quantum memory via Raman scheme based on cold atomic ensembles was performed, in which the Bell's equality was violated in the storage process. In the work of Walmsley's group, the single photon from spontaneously parametric down conversion (SPDC) was stored and retrieved, which showed the heralded single photon quantum memory. However, as they pointed out, there was an inevitable noise from nonlinear four-wave mixing (FWM) process in hot atomic vapour cell. Since the photons encoded in polarization are more easily transported in optical fibre and are robust against decoherence in long-distance transmission channel, it is efficient to realize the quantum interface for connecting different quantum memories based on photonic polarised entanglement, no matter for scalable quantum computation or establishing polarized entanglement distribution in long distance. As said before, Raman quantum memory scheme gives high speed properties in the transmission rate, many people have been attracted in this field. However, realizing Raman quantum memory of photonic polarised entanglement including heralded entangled state or post-selected two-particle entanglement, is a missing and challenge work in the field of quantum information science.

In this paper, we generated heralded single photon through spontaneously Raman scattering process in one atomic ensemble. Then, by using a specific Sagnac interferometer, the heralded photonic hybrid entanglement between path and polarization was stored and retrieved in another atomic ensemble. The measured concurrences before and after storage clearly showed that there was about 20.9% ±7.7% transferring efficiency of quantum entanglement in the process of storage. The obtained visibilities of heralded entangled state before and after storage clearly illustrated the workability of our Raman memory system under heralded quantum regime. In addition, by using two optical interferometers including a actively locking interferometer and the specific Sagnac interferometer, we demonstrated Raman quantum memory of two-particle polarised entangled state, the Bell's equality was violated by 3.2 standard deviations after storage without any noise corrections, the reconstructed density matrices before and after storage were measured and the calculated fidelity between them was 85.0% ±3.4%, showing the entanglement is preserved very well. Our experimental performance of Raman storing heralded single-particle photonic entanglement and two-particle photonic entanglement takes a big step forward to quantum information science.

**Experimental Results**

The media for preparing single photon sources is an optically thick Rubidium (Rb) ensemble, which is trapped in a two-dimensional magneto-optical trap (MOT A) [28]. The spontaneously

Raman scattering (SRS) [29-30] configuration shown in Fig. 1 (a) is the scheme of generating non-classical photon sources in our experiment. The two lasers (pump 1 and pump 2) at wavelength of 780 nm and 795 nm are acted as the input pump lasers. These two lasers are both from an external-cavity diode laser (DL100, Toptica), and the pump 2 is blue-detuned to the atomic transition of $5S_{1/2}(F=3)$->$5P_{1/2}(F'=3)$ with a value of 70 MHz, the pump 1 is resonant with the atomic transition of $5S_{1/2}(F=2)$ ->$5P_{3/2}(F'=3)$. By satisfying the condition of phase matching of FWM, the spontaneously generated Anti-stokes photons (795 nm) and Stokes photons (780 nm) are time-energy entangled, thus resulting in correlated curve in time domain. Such method of preparing photons source is same as Ref. [16], where the quality of photons source is characterized by the factor of α (where α closing to zero describes the nature of generated single photon). By introducing a Mach-Zehnder interferometer with two FWM paths process shown in Fig. 1 (b), the generated photon sources would be entangled in polarised degree of freedom with the active phase stabilization. Through modulating the relative phase of two FWM paths, we could prepare the different photonic polarised entangled states like in [31-32].

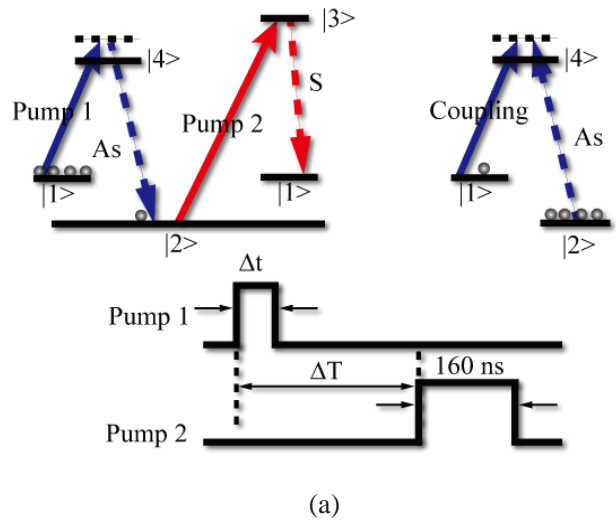

(a)

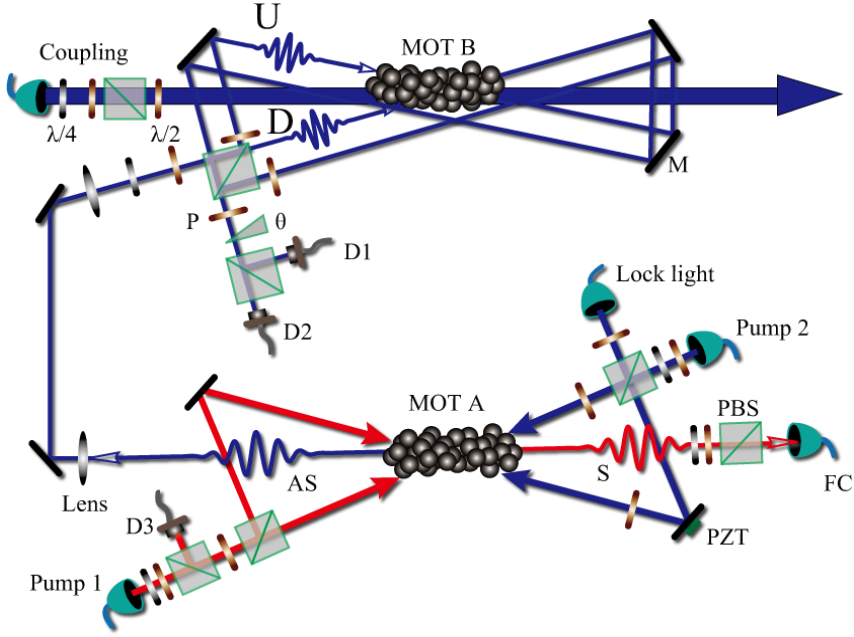

(b)

Fig. 1 (a) The energy levels diagram for our system and simplified time sequence of generating non-classical photon sources. The ground energy levels of |1> and |2> correspond to two metastable levels $5S_{1/2}(F=3)$ and $5S_{1/2}(F=2)$ of Rb 85 atom respectively, |3> and |4> are the excited levels of $5P_{3/2}(F′=3)$ and $5P_{1/2}(F′=3)$ respectively. The pump 1 and pump 2 lasers are modulated by two acoustic optic modulators to be a sequence of $\Delta_t$=25 ns and 160 ns width pulse. MOT: magneto-optical trap; FC: fibre coupler; PBS: polarization beam splitter; λ/2: half-wave plate. λ/4: quarter-wave plate. S: Stokes photon; As: Anti-stokes photon; D1, D2: single photon detector 1 and 2 (PerkinElmer SPCM-AQR-15-FC); D3: home-made photoelectric detector; PZT: Piezoelectric transducer. U, D: up-optical mode and down-optical mode input into MOT B respectively. P: half-wave plate for performing interference. θ: the phase of inserted phase plate.

## 1. Storing entanglement of single photon path and polarization.

Here, we focused on the storage of heralded single photon in another atomic ensemble called MOT B. The single photons used for sequent storage was directly generated from a FWM path in MOT A, another path FWM channel was blocked. The generated Anti-stokes photon was directly input into MOT B with two separated paths U and D by using a specific Sagnac interferometer. These two optical paths were completely covered by atomic cloud in MOT B. A coupling laser with frequency-detuned of +70 MHz was incident into MOT B, and with same angles between these two optical paths of Anti-stokes. Since there was Raman interaction process between atoms and coupling laser, a virtual energy level at a detuning of +70 MHz was excited. The excited virtual energy level coupled the coupling laser and Anti-stokes photon with two ground states |1> and |2> via Λ atomic configuration. The power of coupling laser was 22 mW with a waist of 2 mm, corresponding to the Rabi frequency of 7.8 Γ (Γ is the decay rate of level |4>). The Raman storage should work under bandwidth matching between the memory in MOT B and the Anti-stokes photons (see Method section). We adiabatically turned off and on the coupling laser to realize storage of Anti-stokes photon. By using the specific Sagnac interferometer [33], the single photon's entangled state was prepared as:

$$|\psi_1\rangle = \frac{1}{\sqrt{2}}(|U\rangle|H\rangle + e^{i\theta_1}|D\rangle|V\rangle) \qquad (1)$$

where, |U>, |D> are the states distinguishing the two optical paths (up and down optical modes) in Sagnac interferometer. |H>, |V> represents the horizontal and vertical photonic polarization degree of freedom. Actually, Eq. (1) describes a hybrid entangled state of photonic path and polarization. If we trace the path degree of freedom by detecting the output signal after the PBS in Sagnac interferometer, we would obtain the arbitrary superposition of photonic polarization. Also, if we trace the polarization information of entanglement of |ψ$_1$>, the rest of which is the path information.

In our experiment, we set the relative phase $\theta_1$ to be zero (see Method section). In order to verify the entangled properties of |ψ$_1$> before and after storage, we described this property by using the reduced density matrix ρ in the basis of |$n_U$, $m_D$> with {n,m}={0,1}, which was introduced in Ref. [6], the density matrix was written as:

$$\rho = \frac{1}{P}\begin{pmatrix} p_{00} & 0 & 0 & 0 \\ 0 & p_{10} & d & 0 \\ 0 & d^* & p_{01} & 0 \\ 0 & 0 & 0 & p_{11} \end{pmatrix} \qquad (2)$$

Here, $p_{ij}$ is the probability of finding $i$ photons in mode $U_k$ and $j$ in mode $D_k$, $k$ represents the input or output modes; $d = V(p_{01}+p_{10})/2$ is the coherence between |1$_U$0$_D$> and |0$_U$1$_D$>. $P = p_{00}+p_{10}+p_{10}+p_{11}$.

$V$ is the visibility for interference between modes $U$ and $D$. The visibility could be measured by recording the coincidence counts for detectors 1 and 2 with different phases $\theta$ by rotating the half-wave plate to be 22.5°. The entangled property of density matrix $\rho$ is characterized by the concurrence $C = \frac{1}{P}\max(0, 2|d| - 2\sqrt{p_{00}p_{11}})$. The value of concurrence between 0~1 represents the separate state toward maximally entangled state. In our experiment, the measured concurrences of density matrices before and after storage were $C_{\text{input}}=(5.8\pm0.2)\times10^{-3}$ and $C_{\text{output}}=(1.2\pm0.4)\times10^{-3}$ respectively. In this process, the coupling efficiency and transmission loss were considered, including 30% transmission rate of three home-made Fabry–Perot (F-P) cavity filters and 50% coupling efficiency of fiber. We obtained the factor $\eta = C_{\text{output}}/C_{\text{input}} \sim 20.9\% \pm 7.7\%$, which represents the transfer efficiency of heralded entanglement in the storage process. We obtained the visibilities of input and output state, the measured values of $V_{\text{input}}=86.9\%\pm3.1\%$ before storage and $V_{\text{output}}=82.2\%\pm5.7\%$ after storage were all larger than the benchmark of 70.7%, which clearly shows that the entanglement properties was conserved in the storage process. The contrast factor $\beta = V_{\text{output}}/V_{\text{input}} = 95\pm10\%$ was so high, representing the high fidelity of our memory system.

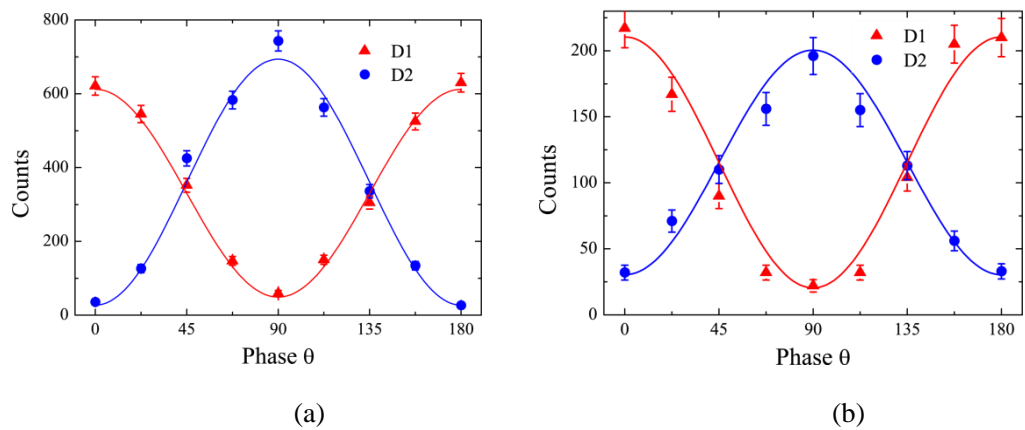

(a)          (b)

Fig. 2 (a) is the coincidence between the Stokes photon and detector 1 (circular data), detector 2 (triangular data) with different phase before storage. (a) Records the coincidence between the Stokes photon and detector 1 (circular data), detector 2 (triangular data) with different phase after storage. In this process, the half-wave plate P was rotated to be 22.5 degree. All experimental data were raw without error corrections. Error bar is ±1 standard deviation.

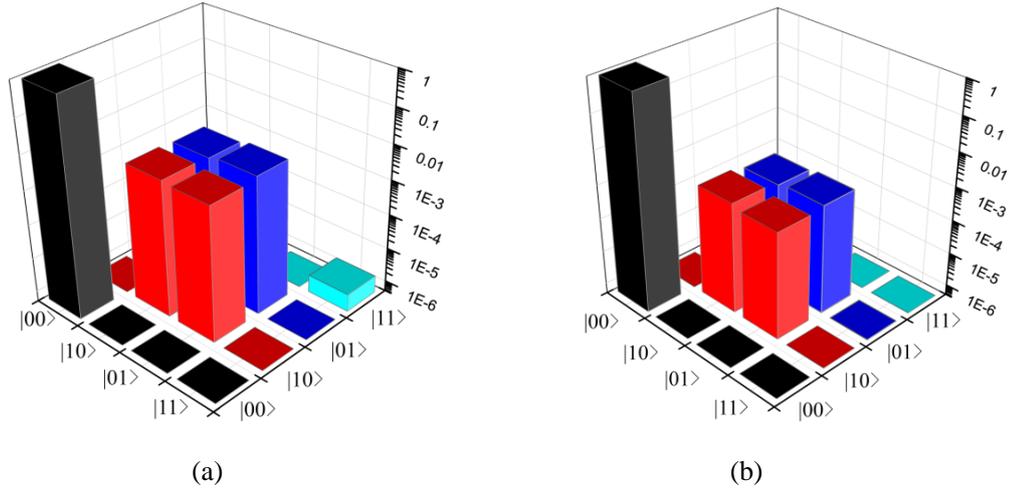

(a)                          (b)

Fig. 3 The density matrices of input state before storage (a) and output state after storage (b). All experimental data were raw without any error corrections.

Table 1. The measurement of $\bar{p}_{ij}$ and concurrences $\bar{C}$ before and after storage.

|  | $\bar{\rho}_{\text{input}}$ | $\bar{\rho}_{\text{output}}$ |
|---|---|---|
| $\bar{p}_{00}$ | $0.990393 \pm 0.00006$ | $0.998166 \pm 0.000008$ |
| $\bar{p}_{10}$ | $(4.59 \pm 0.03) \times 10^{-3}$ | $(9.64 \pm 0.04) \times 10^{-4}$ |
| $\bar{p}_{01}$ | $(5.04 \pm 0.03) \times 10^{-3}$ | $(8.71 \pm 0.04) \times 10^{-4}$ |
| $\bar{p}_{11}$ | $(1.6 \pm 0.2) \times 10^{-6}$ | $(5 \pm 5) \times 10^{-8}$ |
| $\bar{C}$ | $(5.8 \pm 0.2) \times 10^{-3}$ | $(1.2 \pm 0.4) \times 10^{-3}$ |

In this process, the entanglement between photonic path and polarization was stored and retrieved, giving the ability of storing hybrid entangled state. To some extent, the obtained visibility was greater than the threshold of 70.7% of benchmark of Bell's equality [34], showing the workability under quantum regime. The stored Anti-stokes photon encoded in entangled state was triggered by the Stokes photon, thus our storage of single photon's entanglement in polarization and path could be considered as a heralded single photon's entanglement quantum memory. In quantum communication, this heralded Raman memory process for single photon meets the criteria of the DLCZ protocol [35], where entangled memories should be heralded by the detection of a single-photon. This is the basic requirement for establishing scalable quantum networks in future.

## 2. Storing entanglement of two photonic polarizations.

Next, we would store two photonic polarized entanglement based on two cold atomic ensembles. In the experiment, we built up the polarized entanglement between the Anti-Stokes photon and the collective spin excited state of the atomic ensemble by SRS process in MOT A firstly. Then we input the Anti-stokes photon into MOT B for storage. Thus, the entanglement was established between the collective spin excited state of atoms in MOT A and B, it was $|\psi_{aa}\rangle = \frac{1}{\sqrt{2}}(|H\rangle_A|V\rangle_B + e^{i\theta_2}|V\rangle_A|H\rangle_B)$, $\theta_2$ was set to be zero in our experiment. After we retrieved collective spin excited state of atoms in MOT B as Anti-stokes photons, we read the collective spin excited state of the atomic ensemble in MOT A out as Stokes photons. In this process, the storage time in MOT B should be less than in MOT A, made sure storing entanglement (see Method section). By projecting these two retrieved Anti-stokes and Stokes photons into four basis of $|H\rangle$, $|V\rangle$, $(|H\rangle+|V\rangle)/2^{1/2}$, $(|H\rangle-i|V\rangle)/2^{1/2}$), where $H$ and $V$ stands for horizontal and vertical linearly polarisations respectively, we could reconstruct the density matrices of atomic collective spin excitation entangled state by reconstructing the retrieved photonic polarized entangled state.

The prepared photonic polarized entangled state could be written as:

$$|\psi_2\rangle = \frac{1}{\sqrt{2}}(|H\rangle|V\rangle + |V\rangle|H\rangle) \qquad (2)$$

The reconstructed density matrix of prepared entangled state was shown in Fig. 4. The real and imaginary reconstructed density matrices for input state were given in Fig. 4 (a) and (b). By using the formula of $F_1 = Tr(\sqrt{\sqrt{\rho_{input}}\rho_{ideal}\sqrt{\rho_{input}}})^2$, we calculated the fidelity of the reconstructed density matrix by comparing it with the ideal density matrix, which was of 89.3% ±1.7%. After programed storage time, the retrieved density matrices of real part and imaginary part were given in Fig. 4 (c) and (d). Through calculating the fidelity $F_2 = Tr(\sqrt{\sqrt{\rho_{output}} \rho_{input} \sqrt{\rho_{output}}})^2$ with the obtained input density matrix, the obtained fidelity was 85.0% ±3.4%.

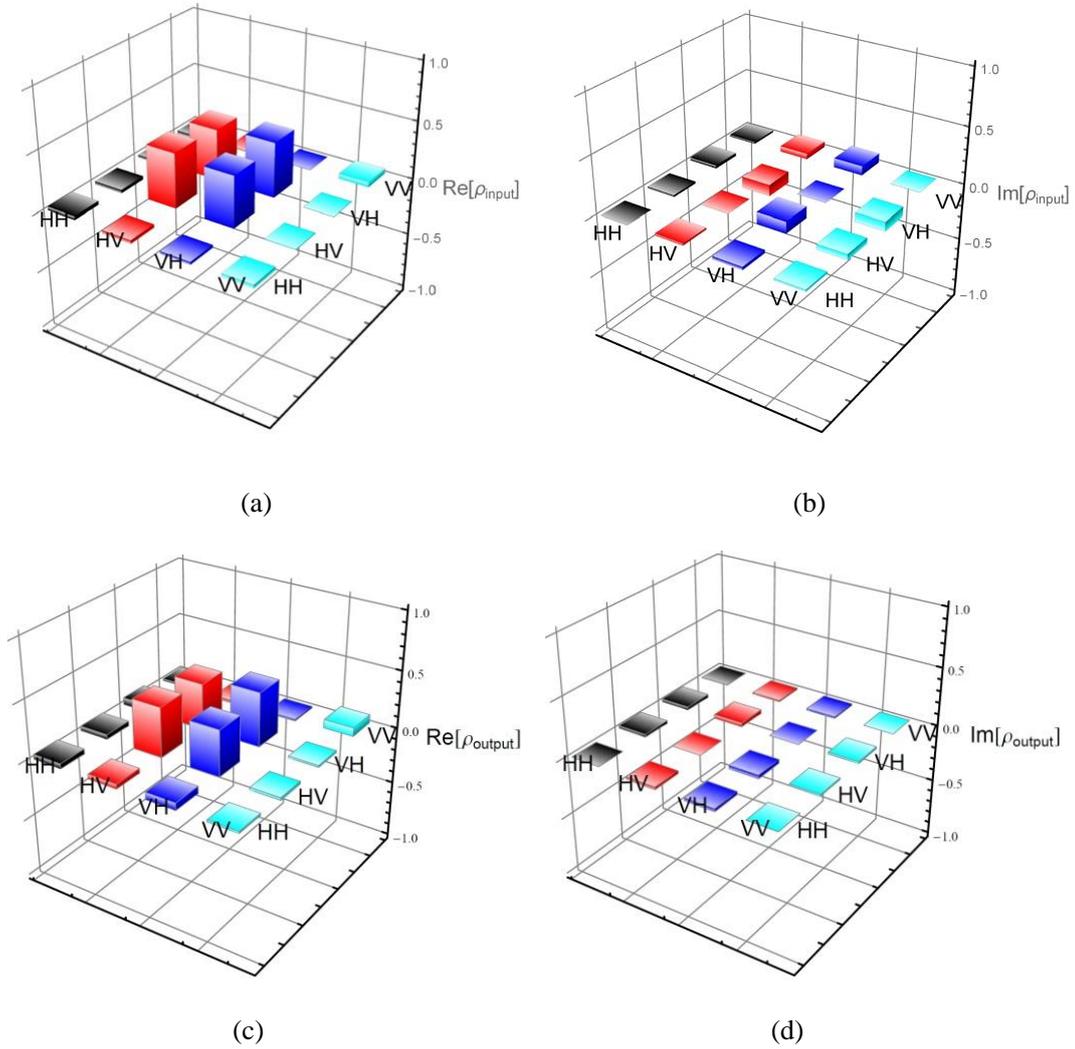

(a)　　　　　　　　　　　　　　　(b)

(c)　　　　　　　　　　　　　　　(d)

Fig. 4. The reconstructed density matrices for input entanglement and output entanglement. (a) and (b) are the reconstructed real and imaginary parts of input density matrix. (c) and (d) are the reconstructed real and imaginary parts of output density matrix. All experimental data are raw without error corrections.

We further characterized the properties of entanglement before and after storage through checking the violation of Bell's inequality, which was the symmetrized version called CHSH inequality [34]. The S is defined as:

$$S = \left| E(\theta_A, \theta_S) - E(\theta_A, \theta_S') + E(\theta_A', \theta_S) + E(\theta_A', \theta_S') \right|$$

Where $\theta_A$, $\theta_S$ were the angles of the phase. $E(\theta_A, \theta_S)$ can be calculated from the coincidence rates at particular orientations,

$$E(\theta_A, \theta_S) = \frac{C(\theta_A, \theta_S) + C(\theta_A + \frac{\pi}{2}, \theta_S + \frac{\pi}{2}) - C(\theta_A + \frac{\pi}{2}, \theta_S) - C(\theta_A, \theta_S + \frac{\pi}{2})}{C(\theta_A, \theta_S) + C(\theta_A + \frac{\pi}{2}, \theta_S + \frac{\pi}{2}) + C(\theta_A + \frac{\pi}{2}, \theta_S) + C(\theta_A, \theta_S + \frac{\pi}{2})}$$

In our experiment, $\theta_A=0$, $\theta_S=\pi/8$, $\theta_A'=\pi/4$, $\theta_S'=3\pi/8$. The calculated S values are 2.40±0.04 for entanglement before storage, after retrieving process, the S value became 2.26±0.08. We also measured the visibilities before and after storage. The results were shown in Fig. 5. The visibilities we calculated were 85.9%±3.0% before and 80.6%±3.5% after memory respectively, which beaten the threshold of 70.7% of benchmark of Bell's equality, showing the preservation under quantum memory process.

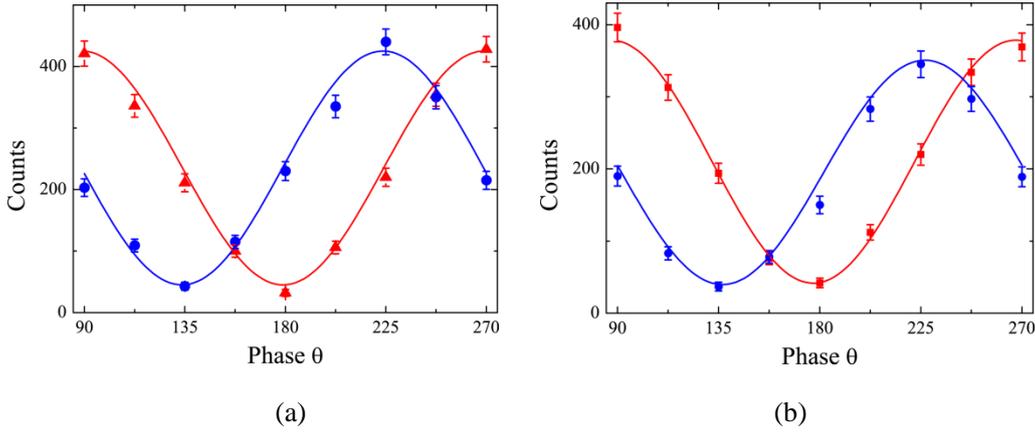

(a)           (b)

Figure.5. The measured interference of retrieved Stokes photons. The red (blue) curve represents the correlated coincidence rate with the Stokes projection of $|H>$ $((|H>-|V>)/2^{1/2})$. (a) was the interference before storage, (b) was the interference curve after retrieval. The data were all raw, without any corrections. Error bar is ±1 standard deviation.

The measured fidelity, visibility and S values clearly showed our two quantum memories with 1 metre distance were established to be polarization-entangled by using Raman quantum memory protocol. As said in beginning, this progress would be essentially significant due to that the different memories in long distance could be established entanglement by using Raman quantum memory [14]. In addition, employing Raman quantum memories to synchronize probabilistic and independent events such as the creation of photon pairs subjected to transmission loss was important, which could fixes the smallest temporal width of the photons. In this way, operating such smallest temporal width of the photons directly affects the overall clock rate of linear optical circuit in quantum computation [36].

**Discussions**

In this paper, in order to perform Raman quantum memory of photonic polarised entanglement, we had stored two type entangled states: 1. Single photon entangled state encoded in photonic polarization and path. 2. Two photon entangled state based on photonic polarization freedom of degree. These two entangled states quantum memories were different type quantum memories in following reasons: the first was based on single-particle system, which was analyzed before, was heralded quantum storage of single photon's entangled state; the second memory process was based on two-particle entangled system, which was in principle a post-selected entanglement quantum memory. If we want to perform a heralded quantum memory of two particle entangled state without any post-selected operations, we need to prepare genius tripod photon entangled states as performing in Ref. [37, 38, 39], where the heralded Bell's state was prepared. However, the quantum memory experiment by using heralded Bell's state is very difficult to achieve under the current experimental conditions, because it needs very long time to record weak tripod coincidence event. Therefore, for establishing heralded long-distance entangled memories distribution, it needs to be explored further.

We also gave the advantages of Raman memory of broad band or high-speed property. In ref. 27, the GHz band width of single photons were stored (the full width at half maximum (FWHM) of photon was 1 ns pulse width). The FWHM of single photon being stored in our memory was ~7 ns (see the supplement material), whereas corresponding to ~140 MHz (0.14 GHz) frequency band width. The second-order cross-correlation $g_{AS,S}(\tau)$ for retrieved signal was measured as 13.6 (see the supplement material), greater than 2 describing the quantum nature [33]. Therefore, our quantum memory showed the ability of storing single photons under the nanosecond level in quantum regime. In principle, the higher broad band photons could be stored by increasing the optical depth (OD). The high OD of 1600 in ref. 14 was used, instead a lower OD~20 in our system was applied. In addition, our system satisfied the condition of far off resonance, which was clearly different from EIT scheme. In Ref. 14, the several GHz detuning was applied, there existed about 500 MHz Doppler line width for theirs hot atomic memory system, thus, the ratio was calculated about 18 for meeting the requirement of far off resonance. In our system, the single photon's detuning could be 200 MHz under our experimental conditions (see the supplement material). In this case, the ratio was calculated as 34.5 by considering the absorption bandwidth of 5.7 MHz for cold atoms, which also met the criteria of far off resonance.

Another important thing we wanted to point out was that the noise from Raman quantum memory for single photons was mainly from FWM process clearly shown in ref. 27, where the hot vapor cell was used. However, in our system, the conditions of a lower OD and cold atoms were used, the FWM noise hadn't been observed, showing the workability under quantum regime. The main noises in our system were from scattering light from coupling laser, which could be reduced to the dark coincidences level by adding three home-made F-P cavity filters with an extinction ratio of $10^7$:1. In principle, it was important to find a most suitable experimental condition for realizing higher speed storage for our system.

**Conclusion**

In conclusion, we experimentally demonstrated two Raman storages of photonic polarised entanglement based on single-particle and two-particle systems respectively. The measured experimental data including concurrences, visibilities and fidelity before and after storage clearly showed the successful Raman storage of true single photons under quantum regime, no matter for heralded entanglement or post-selected entanglement quantum model. The results we obtained were in principle significantly important in the field of quantum information science, more experiments about Raman quantum memory may be explored further to realize a high-speed, long-time, and high-fidelity storage in quantum regime. Our results pave a significant step towards quantum communications based on Raman quantum memory.

## Methods

**Stabilizations of two interferometers.** In our system, we needed to construct two interferometers. One was Sagnac interferometer, was a phase-insensitive interferometer which combined two different path modes ($U$ and $D$). We only needed to manually adjust one of mirrors to obtain the perfect interference we wanted. Another was the phase-sensitive interferometer, a PZT is needed to stabilize relative phase of two FWM paths with optoelectronic feedback. In principle, through adjusting the PZT in this interferometer, we could arbitrarily obtain the four Bell polarised entangled states of prepared Anti-stokes and Stokes photons. The techniques of interferometer for generating entangled state are similar to ref. 32, where the two pump lasers were locked instead of locking two paths of probe optical route in ref. 31. In practically demonstration, the lock light was not collinear with the pump 1 and 2 lasers for reducing the scattering noise of the lock light.

**Bandwidth matching between the Anti-stokes photons and the memory.** Whether the bandwidth of Anti-stokes photons and the memory in MOT B matched or not was affected by the frequency of the coupling and the pump 1 lasers, which could be modulated by two acoustic optic modulators. As described before, the pump 1 laser was input into the MOT A and the coupling laser was input into the MOT B. The atoms in MOT A and MOT B were populated in the atomic ground level of |2>. The Raman excitation absorption spectrum of atomic cloud in MOT B was modulated by changing the frequency of coupling laser. We made the Anti-stokes photons be absorbed completely by the atoms in MOT B due to the atomic Raman transition between the excited virtual energy level and ground state level |2>. Thus, the Raman scattering Anti-stokes photons from MOT A were just right landed in the middle of the spectrum of absorption spectrum in MOT B. Through this method, the bandwidth of Anti-stokes photons and the memory matched well. The maximum efficiency of Raman quantum memory in our experiment was obtained as 26.7%. The efficiency of Raman quantum memory is mainly limited to reabsorption in the forward direction and therefore the maximum efficiency that one can achieve is 54% [40]. This limitation may be solved by using a backward direction.

**Coherence time vs storage time.** In our experiment, the residual geomagnetic field was reduced by using three groups of coils. In MOT A, the generation of Anti-stokes and Stokes photons were separated in time sequence. By using the optical interferometer, the photonic polarised entangled state could be prepared through time-delayed SRS process. In such way, before we read the spin wave out of MOT A as Stokes photon, the Anti-stokes photon was input into MOT B for storing and retrieving. Thus, in order to obtain the perfect signal to noise ratio, we must ensure that the storage time in MOT B was less than the storage time of spin wave in MOT A. In experimental time sequence, the delay time was 160 ns and the storage time we set was 100 ns for simplified demonstration. Theoretically, the storage time could be longer by increasing the coherence time of spin wave in MOT A. The long coherence time of prepared spin wave could be enhanced by using some techniques such as: populating atoms into the magnetic-insensitive level [41] or reducing the angle between the coupling laser and the probe light for enhancing the coherence of spin wave [42-43]. Increasing the storage time of Anti-stokes photon in MOT B also needs the above techniques [41-43], especially for longer time storage the dynamic decoupling method should be introduced [44]. There should be more things to explore for long-time quantum memory of photonic polarized entangled state.

**Error estimation by using Monte Carlo simulations.** All error bars in our experiment were estimated from Poisson statistics and using Monte Carlo simulations. The details are the following: we assume experimental data follows the Poisson distribution. For each data point, we assume 20 random numbers by using Mathematica software, which are in the Poisson distribution. The errors of these quantities are then estimated by the square root of the variance of the 20 values.

**Author contributions**

BSS conceived the idea and experiment. WZ and DSD designed and carried out the experiment with assistance from ZYZ and SS. Data analysis was carried out by DSD and WZ. DSD wrote this manuscript. BSS and GCG supervised the project.

**Acknowledgements**

This work was supported by the National Fundamental Research Program of China (Grant No. 2011CBA00200), the National Natural Science Foundation of China (Grant Nos. 11174271, 61275115, 61435011), the Youth Innovation Fund from USTC (Grant No. ZC 9850320804), and the Innovation Fund from CAS.

# Supplementary Material for

# Raman Quantum Memory of Photonic Polarised Entanglement

**Dong-Sheng Ding, Wei Zhang, Zhi-Yuan Zhou, Shuai Shi, Bao-Sen Shi and Guang-Can Guo**

In this supplement, we gave storage of nanosecond level short timing pulse. And we demonstrated the Raman quantum memory under far off resonance in our system.

**Storing nanosecond level short timing pulse**

In this section, we had tried to obtain the maximum storage bandwidth of our system. We changed the pulse width of pump 1 laser by using an arbitrary function generator AFG 3252, then the pulse width of generated Anti-stokes photons was changed. We reduced the width of Anti-stokes photons as possible as we can. Because time sequence was controlled by timing software of PCI 6602, AOM and AFG 3252, thus the final full width at half maximum of generated Anti-Stokes photons was about ~7 ns under the limit of our system, which was shown by Fig. S1. We performed the Raman quantum memory of these prepared Anti-Stokes photons. The results were given in Fig. S2, Fig. S2(a) was the input signal without the coupling laser and the atoms in MOT B, Fig. S2(b) was the storage data and Fig. S2(c) was the recoded coincidence with Anti-stokes photons blocking but with coupling laser and the atoms in MOT B opening. The efficiency of storage was obtained as 10.3%. The second-order cross-correlation $g_{s1,s2}(\tau)$ for retrieved signal was measured as 13.6, given by Fig. S2(b). In the results, the practical timing pulse width of measured cross-correlated function was larger than the timing pulse width of Anti-stokes photons, because the timing width of cross-correlated function was restricted by the life-time of atomic level |4>. To some extent, the stronger Rabi frequency of pump 2 laser could be used to reduce the time pulse width of cross-correlated function [1,2,3]. If the resolution of our timing controller could be adjusted to be smaller, our system could be performed with lower timing pulse width (< 7ns) in quantum regime.

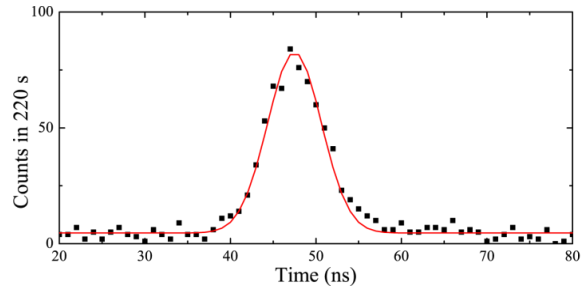

Fig. S1. Measurement of the timing pulse width of Anti-stokes photons excited by pump 1 laser. This measurement is the coincidence between the trigger from AFG 3252 and the detection events of Anti-stokes photons. The red line is the fitted curve by using a formula of $y=y_0+A \exp[-2((t-t_c)/w)^2]$, where $w=6.3$; $y_0=4.6$; $t_c=47.5$; $A=492.9$.

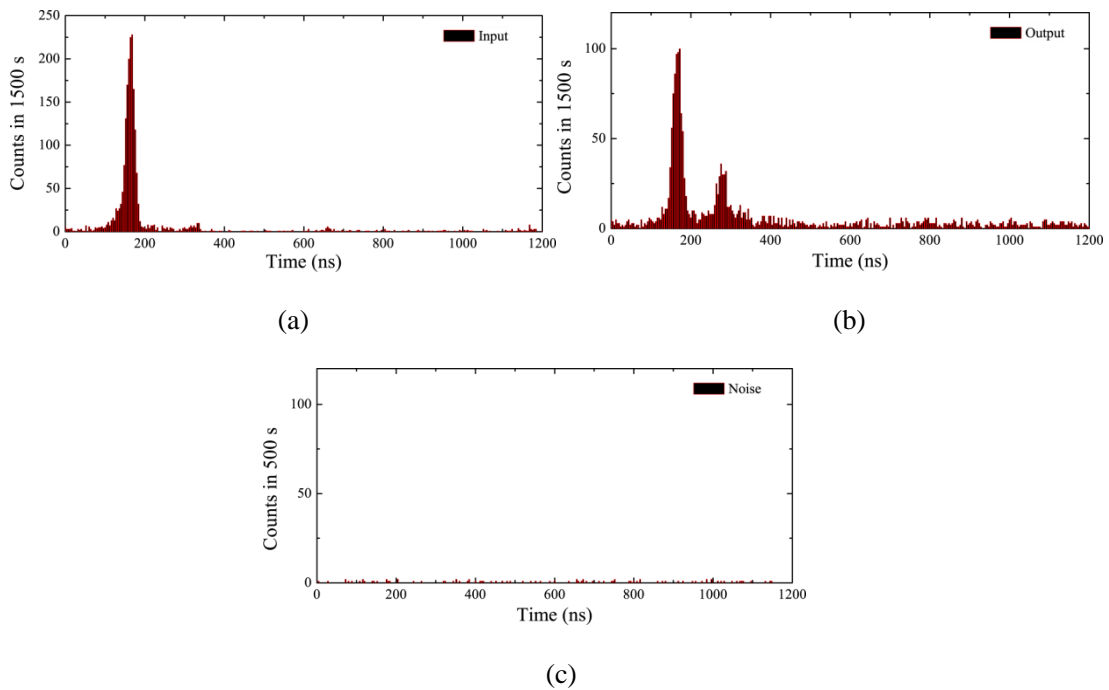

Fig. S2. (a) Coincidence between Anti-Stokes photons and Stokes photons without any storage process. (b) Coincidence between Stokes photons and retrieved Anti-stokes photons. (c) is the coincidence for noise.

**Under far off resonance condition**

In this section, the Raman quantum memory under far off resonance condition was performed in our system. We changed frequencies of the pump 1 and coupling lasers and made them far-detuned +200 MHz with atomic transition $|1\rangle\rightarrow|4\rangle$. Thus, the generated Anti-stokes photons was far-detuned +200 MHz with atomic transition $|2\rangle\rightarrow|4\rangle$. We increased the power of coupling laser to maximum as possible as we can in our laboratory and performed the Raman quantum memory at single-photon level. The atomic absorption bandwidth was about 5.8 MHz considering almost negligible Doppler linewidth at 100 μK. The detuning used in our experiment was 200 MHz, corresponding to the ratio of 34.5, therefore met the far detuning condition. In this process, the width of pump 1 laser was 50 ns, the power of coupling laser was 110 mW with a beam waist of 2 mm, corresponding to Rabi frequency of 17.6 Γ (Γ is the decay rate of level $|4\rangle$). At the same time, we added a home-made F-P cavity filter in our filtering system (the previous number of filters was three), thus the final extinction ratio was about $10^9$:1 for reducing the coupling scattering noise again. The results were given in Fig. S3, Fig. S3(a) was the coincidence for the input signal without the coupling laser and the atoms in MOT B, Fig. S3(b) was the storage data and Fig. S3(c) was the recoded coincidence with Anti-stokes photons blocking but with coupling laser and the atoms in MOT B opening. The measured $g_{AS,S}(\tau)$ for retrieved signal was 5.6. If there were more filters for reducing the Raman scattering noise, the signal to noise ratio would be better. There existed noise in Fig. S3(b), because there was little population in atomic state $|2\rangle$ excited by the Anti-stokes photons.

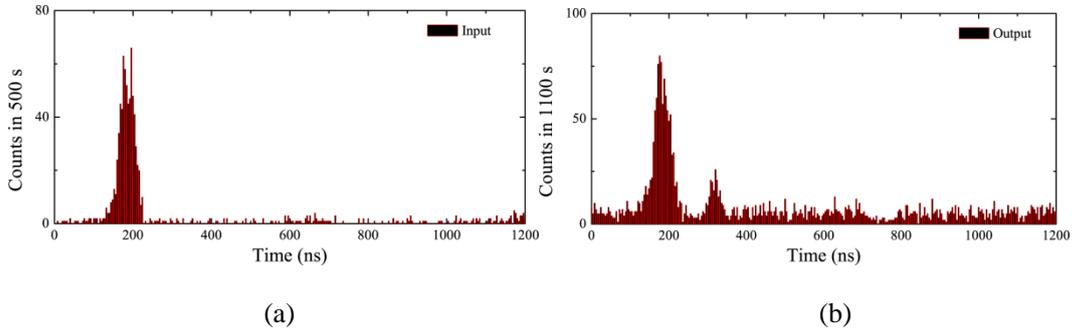

(a)                                    (b)

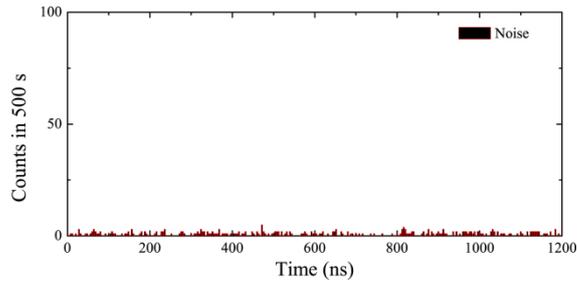

(c)

Fig. S3. Coincidence between the Anti-stokes photons and the Stokes photons under the single photon's detuning of +200 MHz. (a) is the input signal. (b) is the storage of single photons. (c) the recorded noise.